# Mechanism and Improvement of the Harten–Lax–van Leer Scheme for All-Speed Flows

Xue-song Li [*], Chun-wei Gu

*Key Laboratory for Thermal Science and Power Engineering of Ministry of Education, Department of Thermal Engineering, Tsinghua University, Beijing 100084, PR China*

**Abstract:** Based on the three rules developed from the Roe-type scheme, the mechanisms of the classical and preconditioned Harten-Lax-van Leer (HLL) schemes are analyzed. For the classical HLL scheme, the accuracy problem is attributable to the extremely large coefficient of the velocity-derivative term of the momentum equation. For the preconditioned HLL scheme, the global cut-off problem is attributable to the denominator of the coefficients, whereas the particular pressure gradient sensor problem results from the loss of the capability to suppress the checkerboard pressure-velocity decoupling. A new all-speed HLL scheme, which can overcome these problems by only multiplying the momentum-derivative term in the momentum equation by a function related to the local Mach number, is proposed. More importantly, the present study shows the generality of the three rules, which can be powerful tools for analyzing and proposing schemes. The second rule involving the checkerboard problem is also improved by emphasizing that the coefficients of pressure-derivative term in the continuity and energy equations should be changed simultaneously.

**Key words:** Harten-Lax-van Leer scheme, all-speed flows, preconditioning, checkerboard pressure-velocity decoupling, global cut-off problem

## 1. Introduction

---
[*] Corresponding author. Tel.: 0086-10-62794617; fax: 0086-10-62795946
E-mail address: xs-li@mail.tsinghua.edu.cn (X.-S. Li).

The Harten–Lax–van Leer (HLL) scheme [1] is a classical shock-capturing scheme, the family of which includes Rusanov [2], HLLEM [3], HLLE+ [4], HLLC [5] schemes, etc. The HLL-family scheme is one of the most important and popular series of Godunov-type schemes. Similar to other shock-capturing schemes [6, 7], the classical HLL-type scheme is also affected by the highly dissipative behavior in low-Mach number flows. Preconditioning was developed to remedy this problem over the past years, and subsequently, the preconditioned HLL [8] and preconditioned HLLC [9] schemes were proposed. However, the preconditioned HLL scheme suffers not only from the general limitation of preconditioning, which is the global velocity cut-off, but also from a particular limitation of pressure-gradient cut-off. Thus, further research on the HLL scheme for low-Mach number flows should be conducted.

The preconditioning technology has always been promoted by the preconditioned Roe scheme [6, 10, 11]. In recent years, some improvements in the preconditioned Roe scheme were made, and a new concept of the all-speed Roe scheme was developed, where "all-speed" implies overcoming the limitation of the global cut-off problem. For example, Ref. [12] proposed a preconditioned dual-time procedure for an unsteady flow field with cavitations, based on primitive variable pressure, velocity, and enthalpy. Following the idea of preconditioning, Ref. [13–16] proposed an all-speed Roe scheme by modifying only the nonlinear eigenvalues. The same effect can also be achieved by modifying the right eigenvector matrix [17] or the velocity jump normal to the cell interface [18–20]. Ref. [21] reviewed these improved Roe-type schemes [10, 13–20] and discovered three underlying mechanisms for all-speed schemes considering the accuracy, checkerboard pressure-velocity decoupling, and global cut-off problems. Although the three mechanisms are based on the Roe-type schemes, they may have a universal application. Applying these mechanisms to the classical HLL scheme and its preconditioned form, the current paper aims to analyze the reason for the limitation, propose a new all-speed HLL scheme, and show the generality of the three rules.

The outline of the current work is as follows: Section 2 presents the governing equations, as well as the classical HLL scheme and its preconditioned version. Section 3 reviews briefly the three rules, and then applies these rules to analyze the mechanism of the HLL-type schemes. Section 4 proposes a new all-speed HLL scheme based on Section 3 and improves the second rule. Section 5 presents the numerical experiments supporting the theoretical analysis and prediction. Finally, Section 6 states the conclusions.

## 2. Governing Equations and the HLL-type Schemes

### 2.1 Governing Equations

For simplicity, the two-dimensional Euler compressible equations are written as

$$\frac{\partial \boldsymbol{Q}}{\partial t} + \frac{\partial \boldsymbol{F}}{\partial x} + \frac{\partial \boldsymbol{G}}{\partial y} = 0, \qquad (1)$$

where $\boldsymbol{Q} = \begin{bmatrix} \rho \\ \rho u \\ \rho v \\ \rho E \end{bmatrix}$ is the vector of conversation variables; $\boldsymbol{F} = \begin{bmatrix} \rho u \\ \rho u^2 + p \\ \rho uv \\ u(\rho E + p) \end{bmatrix}$ and

$$G = \begin{bmatrix} \rho v \\ \rho uv \\ \rho v^2 + p \\ v(\rho E + p) \end{bmatrix}$$ are the vectors of Euler fluxes; $\rho$ is the fluid density; $p$ is the pressure; $E$ is the total energy; and $u, v$ are the velocity components in the Cartesian coordinates $(x, y)$, respectively.

## 2.2 The HLL Scheme

The numerical fluxes of the classical HLL scheme can be expressed as follows:

$$F^{HLL} = \begin{cases} F_L & S_L > 0 \\ F_{hll} & S_L \leq 0 \leq S_R \\ F_R & S_R < 0 \end{cases}, \tag{2}$$

where $F_{hll} = \dfrac{S_R F_L - S_L F_R + S_R S_L (Q_R - Q_L)}{S_R - S_L}$, (3)

and the signal velocities $S_R$ and $S_L$ are defined as:

$$S_L = \min(U_L - c_L, U_R - c_R), \tag{4}$$

$$S_R = \min(U_L + c_L, U_R + c_R), \tag{5}$$

where $U = n_x u + n_y v$ is the velocity normal to the cell interface; $c$ is the sound speed; and $n_x$ and $n_y$ are the components of the face normal vector, respectively.

Notably, $F^{HLL} = F_{hll}$ is for the low-Mach number flows.

## 2.3 The preconditioned HLL Scheme

According to the preconditioning technology, the preconditioned HLL scheme was proposed [8] by redefining the signal velocities:

$$S_L' = \min(U_L' - c_L', U_R' - c_R'), \tag{6}$$

$$S_R' = \min\left(U_L' + c_L', U_R' + c_R'\right), \tag{7}$$

where the pseudo-velocity and pseudo-sound speed are defined as:

$$U' = \frac{1}{2}(1+\theta)U, \tag{8}$$

$$c' = \frac{1}{2}\sqrt{4c^2\theta + (1-\theta)^2 U^2}, \tag{9}$$

$$\theta = \frac{U_r^2}{c^2}. \tag{10}$$

The key is the definition of $U_r$. In theory, the value of $U_r$ should be related to the local velocity as follows:

$$U_r = \min\left[|U|, c\right]. \tag{11}$$

However, in practice, the $U_r$ of the preconditioned HLL scheme is expanded as follows to avoid computational instability:

$$U_r = \min\left[\max\left(|U|, KU_{ref}, \varepsilon\sqrt{\frac{|\Delta p|}{\rho}}\right), c\right]. \tag{12}$$

Aside from the global velocity $U_{ref}$, which is the general limitation of preconditioning, Eq. (12) also suffers from the pressure gradient sensor $|\Delta p|$, which is the particular limitation of the preconditioned HLL scheme. The coefficient $\varepsilon$ is sensitive to the instability and should be carefully adjusted according to the specific problem [8, 22]. The root of the problem will be determined in the next chapter.

## 3. The Analysis of the Mechanism of the HLL-type Schemes

The numerical dissipation of the continuity and momentum equations can always be expressed as the sum of the velocity-derivative and pressure-derivative terms:

$$F_d = c_u \Delta u + c_p \Delta p, \tag{13}$$

where $c_u$ and $c_p$ are the coefficients of the velocity-derivative and pressure-derivative terms, respectively.

Through the analyses of the five Roe-type schemes, Ref. [21] proposed three rules that can be rewritten as follows:

(1) The accuracy problem of the shock-capture scheme is only attributable to $c_u = \mathrm{O}(c)$ of the momentum equation. The key to correcting the accuracy problem is by setting $c_u \leq \mathrm{O}(c^0)$, regardless of the specific method adopted by the improved scheme.

(2) The checkerboard problem is attributable to the order of $c_p$, especially for the continuity equation. The reasonable interval of the order of $c_p$ is $\left[c^{-1}, c^0\right]$. The order of $\mathrm{O}(c^{-1})$ also allows a weak checkerboard. Increasing $c_p$ can suppress the checkerboard better. The order of $\mathrm{O}(c^0)$ may suppress the checkerboard completely, but may result in an error in the continuity equation. The instability may occur as a result of the uncontrolled checkerboard when $c_p < c^{-1}$ or because of the large error in the continuity equation when $c_p > c^0$. Additionally, the effect of the $\Delta p$ term can be replaced by the time-marching momentum interpolation method [16] for greater accuracy.

(3) The global cut-off problem is attributable to the pseudo-velocity and pseudo-sound speed in the denominator of the coefficients. However, the numerator of the coefficients can be treated as separate from the denominator. Thus, the global cut-off is unnecessary.

To apply these three rules and considering

$$\Delta \rho = \frac{\Delta p}{c^2}, \tag{14}$$

the HLL scheme in the low-Mach number limit, Eq. (3), can be rewritten as follows:

$$F^{HLL} = F_c + F_d, \tag{15}$$

$$F_c = \frac{S_R F_L - S_L F_R}{S_R - S_L}, \tag{16}$$

$$F_d = \frac{S_R S_L}{S_R - S_L} \begin{bmatrix} c^{-2}\Delta p \\ \Delta(\rho u) \\ \Delta(\rho v) \\ \Delta(\rho E) \end{bmatrix}. \tag{17}$$

For the classical HLL scheme, $S_R = \mathrm{O}(c)$ and $S_L = \mathrm{O}(c)$, whereas for the preconditioned HLL scheme, $S_R' = \mathrm{O}(c^0)$ and $S_L' = \mathrm{O}(c^0)$. Then, for both schemes,

$$F_c \approx \frac{F_L + F_R}{2}, \tag{18}$$

which does not result in any problem.

However, $F_d$ of the two schemes differ because:

$$\frac{S_R S_L}{S_R - S_L} = \mathrm{O}(c), \tag{19}$$

$$\frac{S_R' S_L'}{S_R' - S_L'} = \mathrm{O}(c^0). \tag{20}$$

According to the first rule, the reason the classical HLL scheme suffers from the accuracy problem is clear, but the preconditioned HLL scheme can remedy this issue.

According to the second rule, however, the preconditioned HLL scheme suffers from the checkerboard problem because it changes the order of the coefficient of $\Delta p$ in the continuity equation from $\mathrm{O}(c^{-1})$ to $\mathrm{O}(c^{-2})$. Consequently, the preconditioned HLL scheme needs the $\Delta p$-type cut-off in Eq. (12). Selecting the coefficient $\varepsilon$ of the $\Delta p$-type cut-off is difficult because of the lack of a sound theoretical background.

According to the third rule, the preconditioned HLL scheme suffers from the global cut-off problem because it treats both the numerator and the denominator of the pseudo-velocity and pseudo-sound speed terms as equal.

## 4. New All-Speed HLL Schemes

According to the three rules, novel and improved schemes for all-speed flows can be easily proposed. For example,

$$F^{A-HLL} = \frac{S_R F_L - S_L F_R}{S_R - S_L} + \frac{S_R S_L}{S_R - S_L} \begin{bmatrix} \gamma_1 \Delta \rho \\ f(M) \Delta(\rho u) \\ f(M) \Delta(\rho v) \\ \gamma_2 \Delta(\rho E) \end{bmatrix}, \quad (21)$$

where $\gamma_1 = 1$, $\gamma_2 = 1$, and the function $f(M)$ can only be related to the local Mach number, with the simplest version as:

$$f(M) = \min(M, 1). \quad (22)$$

Such a function indicates that the new scheme is an all-speed scheme because it does not need the global cut-off.

Compared with the classical HLL scheme, the only modification required is the multiplication of the momentum-derivative term in the momentum equation by the function $f(M)$. The behavior of the new all-speed scheme may be determined as follows:

(1) According to the first rule, the modification is satisfactory for accuracy, and the new scheme has proper numerical dissipation;

(2) According to the second rule, the new scheme is stable, with a weak checkerboard because the coefficient of $\Delta p$ in the continuity equation remains in the order of $\text{O}(c^{-1})$; and

(3) According to the third rule, the new scheme can avoid the global cut-off limitation because it changes only the numerator.

The above prediction will be confirmed by the numerical experiments in the following chapter.

Interestingly, the following chapter indicates that increasing $\gamma_1$ cannot suppress the checkerboard better, thus seeming to be inconsistent with the second rule. In fact, the second rule is a semi-empirical viewpoint based on the Roe-type scheme, which uniformly modifies the coefficients of $\Delta p$ terms of all governing equations. More accurately, from the asymptotic analysis [6, 15, 21], the energy equation always provides the same constraint as the continuity equation. This condition indicates that the energy equation should have the same modification as the continuity equation, i.e., $\gamma_2 = \gamma_1$.

## 5. Numerical Experiments

To validate the analysis in Chapter 4, a typical low-Mach number test case, the two-dimensional Euler flow past a cylinder, is performed with an inflow Mach number of 0.01 and 72*100 O-type grid points along the circumference and radius. To discuss the schemes themselves, the first-order accuracy is adopted for the HLL-type schemes. The explicit local time-marching algorithm is used for computation, and the Courant–Friedrichs–Lewy (CFL) number is taken as 0.2 for the first-order accuracy.

Fig. 1 shows that the pressure contour is given as a benchmark by the preconditioned Roe with the second-order reconstruction of Monotone Upstream-centered Schemes for Conservation Laws (MUSCL) because it is very close to the theoretical solution.

Fig. 2 shows that the classical HLL scheme produces a solution resembling the pure viscous Stokes flow for the inviscid flow, indicating that the numerical dissipation of the original version of the shock-capturing scheme is extremely large to obtain the physical solution. Adopting the preconditioned HLL scheme with $\varepsilon = 0$ in Eq. (12), even if the good solution in Fig. 1 is employed as the initial field, the severe checkerboard immediately appears around the solid boundary, as shown in Fig. 3, consequently resulting in computational divergence.

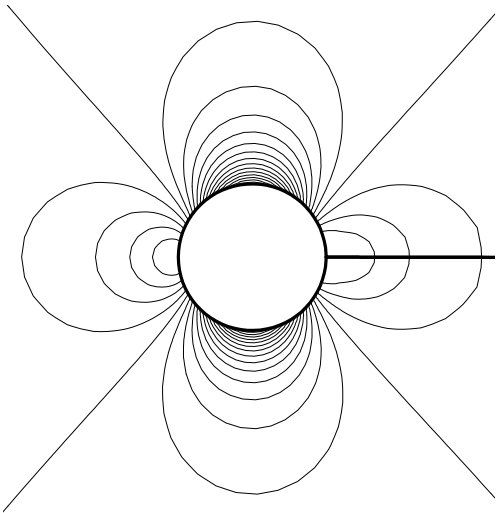 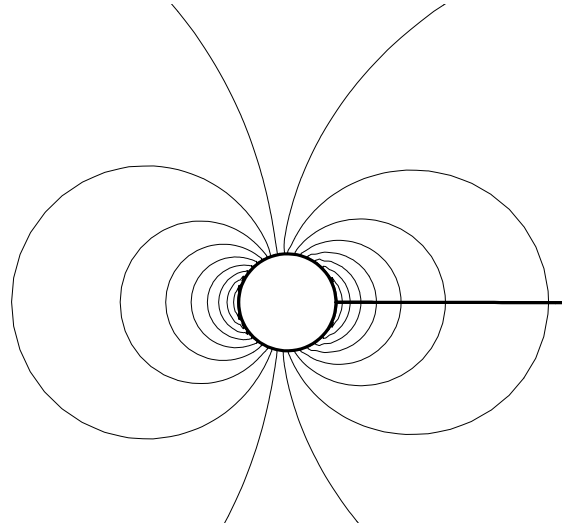

Fig. 1 Second-order preconditioned Roe scheme     Fig. 2 Classical HLL scheme

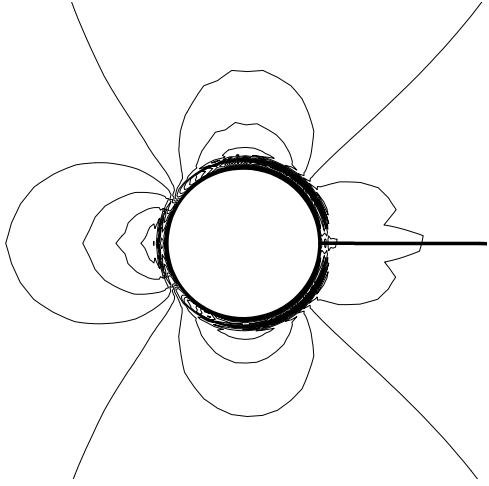 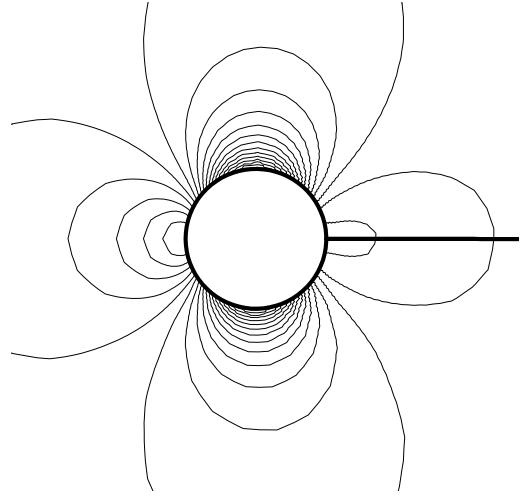

Fig. 3 Preconditioned HLL scheme    Fig. 4 New HLL scheme ($\gamma_1 = \gamma_2 = 1$)

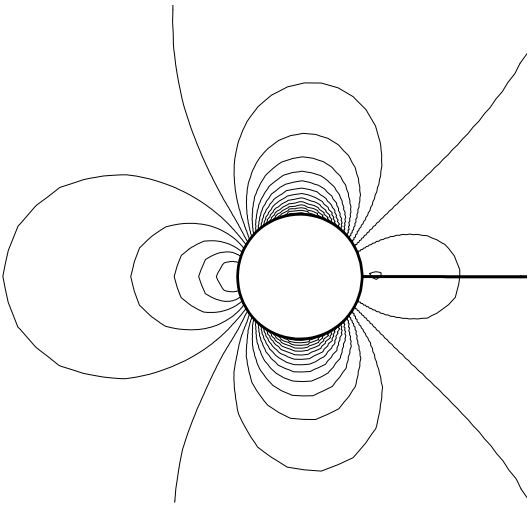 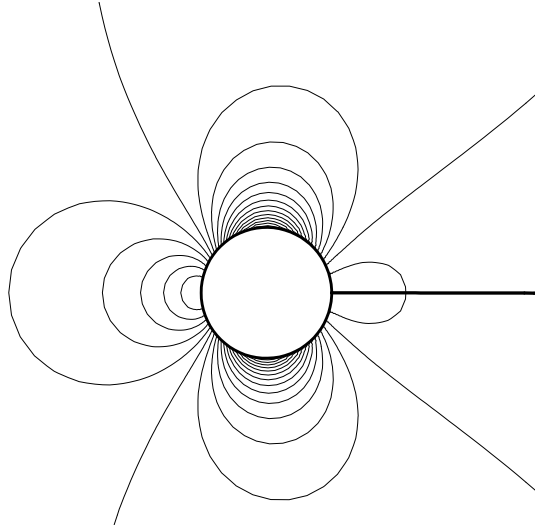

Fig. 5 New HLL scheme ($\gamma_1 = 10, \gamma_2 = 1$)    Fig. 6 New HLL scheme ($\gamma_1 = \gamma_2 = 10$)

As predicted in Chapter 4, the new HLL scheme Eqs. (21) and (22) can produce reasonable results as shown in Figs. 4 to 6. A weak checkerboard is found around the cylinder, but the computation can maintain stability. The divergence of these schemes from that shown in Fig. 1 is due to the first-order accuracy, which can be remedied by the second-order reconstruction. Compared with Fig. 4, where $\gamma_1 = \gamma_2 = 1$, the checkerboard problem does not change markedly in Fig. 5 when only $\gamma_1$ is increased to 10, but can be improved considerably in Fig. 6, when $\gamma_1 = \gamma_2 = 10$. The CFL number should decrease with greater $\gamma_1$ because of the increasing error in the continuity equation.

## 5. Conclusions

Based on the three rules developed from the Roe-type scheme, the mechanisms of the classical and preconditioned HLL schemes were analyzed, and their advantages and disadvantages were explored. Subsequently, a new all-speed HLL scheme, which has a better behavior than the preconditioned HLL scheme, was proposed. Numerical experiments confirm the theoretical analysis and prediction. More importantly, the present study also showed the generality of the three rules, which can be the powerful tools for analyzing the developed scheme and proposing a new and better method.

Likewise, the second rule involving the checkerboard problem is notably semi-empirical, and can be improved through experience. The current study further found that changing the coefficients of $\Delta p$ term in the continuity and energy equations simultaneously should be considered.

## Acknowledgments

The present work is supported by Project 50806037 of the National Natural Science Foundation of China.